# Approaches for Sentiment Analysis on Twitter: A State-of-Art study

Harsh Thakkar and Dhiren Patel

Department of Computer Engineering, National Institute of Technology, Surat-395007, India
{harsh9t,dhiren29p}@gmail.com

**Abstract.** Microbloging is an extremely prevalent broadcast medium amidst the Internet fraternity these days. People share their opinions and sentiments about variety of subjects like products, news, institutions, etc., every day on microbloging websites. Sentiment analysis plays a key role in prediction systems, opinion mining systems, etc. Twitter, one of the microbloging platforms allows a limit of 140 characters to its users. This restriction stimulates users to be very concise about their opinion and twitter an ocean of sentiments to analyze. Twitter also provides developer friendly streaming API for data retrieval purpose allowing the analyst to search real time tweets from various users. In this paper, we discuss the state-of-art of the works which are focused on Twitter, the online social network platform, for sentiment analysis. We survey various lexical, machine learning and hybrid approaches for sentiment analysis on Twitter.

## 1 Introduction

Sentiment analysis is a challenge of the Natural Language Processing (NLP), text analytics and computational linguistics. In a general sense, sentiment analysis determines the opinion regarding the object/subject in discussion. Its initial use was made to analyze sentiment based on long texts such as letters, emails and so on. It is also deployed in the field of pre- as well as post-crime analysis of criminal activities. With the explosion of internet applications such as microbloging websites, forums and social networks, this field of study gained its limelight of late. People discuss, comment and criticize various topics, and write reviews, recommendations, etc. using these applications. User generated data carries a lot of valuable information on products, people, events, and so on. The cut throat competition of the modern world has led to advanced mechanisms of customer feedback and satisfaction-to-innovation cycle. The large user generated content requires the use of automated techniques for mining and analyzing since crowd sourced mining and analysis are difficult. Work in this area originated early from its applications in blogs [1] and product/movie [2] reviews. Today, traditional news outlets have an online version of their news.

Applying this field with the microbloging fraternity is a challenging job. This challenge became our motivation. Substantial research has been carried out in both machine learning and the lexical approaches to sentimental analysis for social networks. We summarize the work done so far will aim to improve the existing approaches Rest of the paper is organized as follows: Section 2 discusses background of sentiment analysis, section 3 summarizes the results obtained from the survey of approaches to sentiment analysis, and finally in section 4 we conclude our paper with conclusion and references at the end.

## 2  Background

Sentiment analysis has been practiced on a variety of topics. For instance, sentiment analysis studies for movie reviews [4], product reviews [5], and news and blogs ([3], [6]). In this section, Twitter specific sentiment analysis approaches are reported. The research on sentiment analysis so far has mainly focused on two things: identifying whether a given textual entity is subjective or objective, and identifying polarity of subjective texts [3]. Most sentiment analysis studies use machine learning approaches.

In sentiment analysis domain, the texts belong to either of positive or negative classes. There may also be multi-valued or binary classes like positive, negative and neutral (or irrelevant). The core complexity of classification of texts in sentiment analysis with respect to that of other topic-based cataloging is due to the non-usability of keywords [2], despite the fact that the number of classes in sentiment analysis is less than that in the later approach by [4].

Opinion mining (sentiment extraction) is employed on Twitter posts by means of following techniques

- Lexical analysis
- Machine learning based analysis
- Hybrid/Combined analysis

### 2.1  Lexical analysis

This technique is governed by the use of a dictionary consisting pre-tagged lexicons. The input text is converted to tokens by the Tokenizer. Every new token encountered is then matched for the lexicon in the dictionary. If there is a positive match, the score is added to the total pool of score for the input text. For instance if "dramatic" is a positive match in the dictionary then the total score of the text is incremented. Otherwise the score is decremented or the word is tagged as negative. Though this technique appears to be amateur in nature, its variants have proved to be worthy ([11], [8]). Fig. 1 shows the working of a lexical technique.

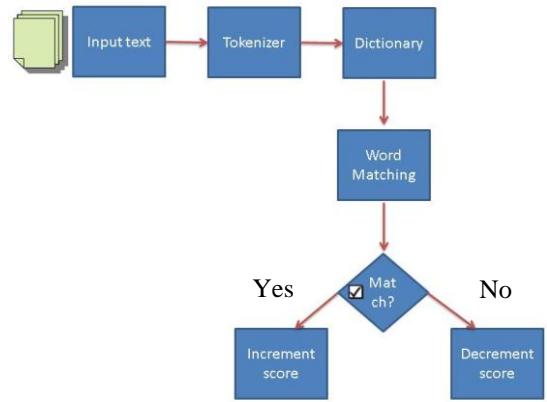

**Fig. 1.** Working of a lexical technique.

The classification of a text depends on the total score it achieves. Considerable amount of work has been devoted for measuring which best lexical information works. An accuracy of about 80% on single phrases can be achieved by the use of hand tagged lexicons comprised of only adjectives, which are crucial for deciding the subjectivity of an evaluative text as demonstrated by [9]. The author of [7] extended this work making use of same methodology and tested a database of movie reviews, reported an accuracy of mere 62%. Other than the hand tagged lexicon approach, [2] came up with a variant by utilizing internet search engine for marking the polarity of words included in work of [7]. They used two AltaVista search engine queries: target word + 'good' and other target word + 'bad'. The score was evaluated by the search that yielded the max number of hits, which reported to improve the earlier accuracy from 62% to 65%. In Subsequent research ([8], [10]) the scoring of words was accomplished by using the WordNet database. They compared the target word with two pivot words ('good' and 'bad') and found the Minimum Path Distance (MPD) between the words in the WordNet pyramid. The MPD is the converted to an incremental score, which is stored in the word dictionary. This variant was reported to yield accuracy [10] of 64%. The author of [11] proposes another method which presents an alternative to ([8], [10]), taking motivation from [9], by evaluating the semantic gap between the words simply subtracting the set of positive ones from the negative ones yielding 82% accuracy. Lexical analysis has a limitation: its performance (in terms of time complexity and accuracy) degrades drastically with the exponential growth of the size of dictionary (number of words).

### 2.2 Machine learning based analysis

Machine learning is one of the most prominent techniques gaining interest of researchers due to its adaptability and accuracy. In sentiment analysis, mostly the su-pervised learning variants of this technique are employed. It comprises of three stages: Data collection, Pre-processing, Training data, Classification and plotting results. In the training data, a collection of tagged corpora is provided. The Classifier is presented a series of feature vectors from the previous data. A model is created based on the training data set which is employed over the new/unseen text for classification purpose. In machine learning technique, the key to accuracy of a classifier is the selection of appropriate features. Generally, unigrams (single word phrases), bi-grams (two consecutive phrases), tri-grams (three consecutive phrases) are selected as feature vectors. There are a variety of proposed features namely number of positive words, number of negative words, length of the document, Support Vector Machines (SVM) ([14], [15]), and Naïve Bayes (NB) algorithm [16]. Accuracy is reported to vary from 63% to 80% depending upon the combination of various features selected.

Fig. 2 shows the typical number of steps involved in a machine learning technique. Working of this technique can be explained as follows:

**First Step:** **Data Collecting** – in this stage data to be analyzed is crawled from various sources like Blogs, Social networks (Twitter, MySpace, etc.) depending upon the area of application.

**Second Step:** **Pre-processing** – In this stage, the acquired data is cleaned and made ready for feeding it into the classifier. Cleaning includes extraction of keywords and symbols. For instance – Emoticons are the smiley used in textual form to represent emotions e.g. ":-)", ":)", "=)", ":D", ":-(", ":(", "=(", ";(", etc.. Correcting the all uppercase and all lowercase to a common case, removing the non-English (or proffered language texts), removing un-necessary white spaces and tabs, etc.

**Third Step:** **Training Data** – A hand-tagged collection of data is prepared by most commonly used crowd-sourcing method. This data is the fuel for the classifier; it will be fed to the algorithm for learning purpose.

**Fourth Step:** **Classification** – This is the heart of the whole technique. Depending upon the requirement of the application SVM or Naïve bayes is deployed for analysis. The classifier (after completing the training) is ready to be deployed to the real time tweets/text for sentiment extraction purpose.

**Fifth Step:** **Results** – Results are plotted based on the type of representation selected i.e. charts, graphs, etc. Performance tuning is done prior to the release of the algorithm.

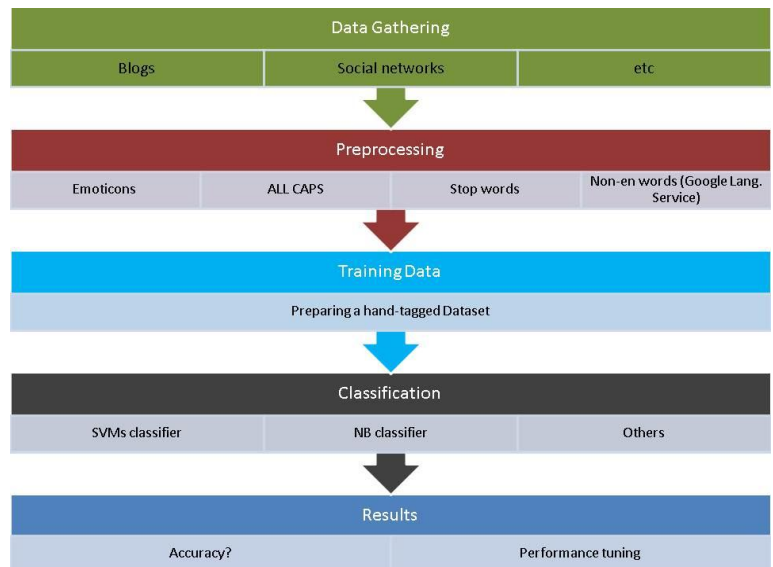

**Fig. 2.** Steps involved in the machine learning approach.

The machine learning technique faces challenges in: designing a classifier, availability of training data, correct interpretation of an unforeseen phrase. It overcomes the limitation of lexical approach of performance degradation and it works well even when the dictionary size grows exponentially.

### 2.3 Hybrid analysis

The advances in sentiment analysis lured researchers to explore the possibility of a hybrid approach which could collectively exhibit the accuracy of a machine learning approach and the speed of lexical approach. In [17] authors use two-word lexicons and an unlabeled data, dividing these two-word lexicons in two discrete classes negative and positive. Pseudo documents encompassing all the words from the set of chosen lexicons are created. Then computed the cosine similarity amongst the pseudo documents and the unlabeled documents. Depending upon the measure of similarity, the documents were either assigned a positive or a negative sentiment. This training dataset was then fed to a naïve bayes classifier for training purpose.

Another approach presented by [18], derived a "unified framework" using background lexical information as word class associations. Authors renewed information for particular areas using the available datasets or training examples and proposed a classifier called as Polling Multinomial Classifier (PMC) (also known as the multinomial naïve bayes). Manually labeled data was incorporated for training purpose. They claimed that making use of lexical knowledge improved performance. Another variant of this approach was presented by [19]. But so far only [18] have been able to claim good results.

## 3 Summary

In a survey conducted by [20], comparison of all approaches has showed that best results have been observed from machine learning approaches, and least by lexical approaches. However, without any proper training of a classifier in machine learning approach results may deteriorate drastically. Work is being carried on hybrid approaches; hence so far only limited information is available to our knowledge.

The techniques were tested by [20] on a movie review & recommendation and news review area based on user tweets. Their results seem to be promising for further research. For the training purpose of the classifiers, they used: Cornel movie review dataset [2], General inquirer adjective lists [12], yahoo web search API [13], Porter stemmer [22], WordNet Java API [23], Stanford log linear POS tagger with Penn Treebank tag set [23], WEKA M.L. java API (only for Machine Learning purpose) [24], SVM-light ML implementation (M.L. classifier) [25]. The results are summarized in Table 1.

**Table 1.** Summary of accuracy in % of various techniques of sentiment analysis derived from tests carried out by [20] and our other survey (* - approx).

| Approach | Including features | Accuracy |
|---|---|---|
| Lexical [20] | Baseline (as is) | 50 |
| Lexical [20] | Baseline, stemming | 50.2 |
| Lexical ([8],[10]) | Baseline , WordNet | 60.4 |
| Lexical [20] | Baseline , Yahoo web search | 57.7 |
| Lexical [20] | Baseline, all above | 55.7 |
| Machine Learning ([14],[15]) | SVM , Unigrams | *~77 |
| Machine Learning ([14],[15]) | SVM , Unigrams , Aggregate | 65 - 68 |
| Machine Learning [16] | Naïve Bayes , Unigrams | 75 - 77 |
| Machine Learning [16] | Naïve bayes , Unigrams , Aggregate | ~77-78 |

On the other hand hybrid approaches are showing the following general sentiment analysis results:

**Table 2.** Summary of accuracy in % of various hybrid approaches derived from tests carried out by [20] and our other survey.

| Approach | Features | Accuracy |
|---|---|---|
| Hybrid [17] | Class-two Naïve bayes , unlabeled data | ~64 |
| Hybrid [19] | SVM + Naïve bayes , emoticons | 70 |
| Hybrid [18] | Class-two Naïve bayes , twitter datasets | ~84 |

## 4 Conclusion

Open social networks are best examples of sociological trust. The exchange of messages, followers and friends and varying sentiments of users provide a crude platform to study behavioral trust in sentiment analysis domain. Machine learning approaches have been so far good in delivering accurate results. Depending upon the application, the success of any approach will vary. Lexical approach is a ready-to-go and doesn't require any prior information or training. While on the other hand machine learning requires a well-designed classifier, huge amount of training data sets and performance tuning prior to deployment. Hybrid approach has so far displayed positive sentiment as far as performance is concerned. Though they have been deployed using unigrams and diagrams, their performance is worse on trigrams. This definitely leaves researchers to explore the terrain.